\documentclass{aa}

\usepackage{graphicx}
\usepackage{txfonts}

\begin{document}

   \title{Delayed maximum energy solar energetic particle events:}

   \subtitle{Statistical analysis from Solar Orbiter}

   \author{R. C. Allen\inst{1}
   \and
   G. C. Ho\inst{1}
   \and
   G. M. Mason\inst{2}
   \and
   Z. Ding\inst{3}
   \and
   M. H. Walker\inst{4}
   \and
   A. Kouloumvakos\inst{2}
   \and
   R. F. Wimmer-Schweingruber\inst{3}
   \and
   J. Rodriguez-Pacheco\inst{5}
   \and
   S. K. Vines\inst{1}
   \and
   R. J. Filwett\inst{6}
   \and
   Z. Xu\inst{8}
   \and
   C. M. S. Cohen\inst{7}
   }

\institute{Southwest Research Institute, San Antonio, TX, USA\\
        \email{robert.allen@swri.org}
    \and
    Johns Hopkins Applied Physics Lab, Laurel, MD, USA
   \and
   Institut f\"{u}r Experimentelle und Angewande Physik, Christian-Albrechts-Universit\"{a}t zu Kiel, Kiel, Germany
   \and
   The Johns Hopkins University, Baltimore, MD, USA
   \and
   Space Research Group, Universidad de Alcal\'{a}, Alcal\'{a} de Henares, Madrid, Spain
   \and
   Montana State University, Bozeman, MT, USA
   \and
   California Institute of Technology, Pasadena, CA, USA
   }

   \date{Received 2 September 2025; accepted 24 November 2025}
 
  \abstract{Investigations of solar energetic particles (SEPs) have long utilized the dispersive nature of onset times, i.e., earlier arrival of higher-energy particles compared to lower-energy particles, to infer information such as the path length to the acceleration site at the time of initial particle release. However, recent observations by Solar Orbiter and Parker Solar Probe have begun to characterize SEP events with an apparent delay in arrival times of the higher-energy portion of the particle distribution, above a critical energy separating the delayed particles from that of the typical velocity dispersion signature at lower energies. Features of these delayed maximum energy (DME) SEP events, sometimes referred to as ``inverse velocity dispersion'' events, may provide new insight into the impacts of magnetic connectivity to locations along an expanding coronal mass ejection (CME)-driven shock wave, variations of acceleration along the shock surface, and transport effects in the inner heliosphere. This study focuses on the occurrence rate and characteristics of DME events observed by Solar Orbiter relative to their footpoint locations with respect to the initial flare site. These DME events show a bias in occurrence rate towards events when the observer's footpoints were westward of the associated flare location. Additionally, estimated locations at which the highest-energy particles of DME events are released into the flux tube suggest continued release of increasingly higher-energy particles from the CME-driven shock into the connected flux tube well into the inner heliosphere. This indicates that DME events are due to inner heliospheric effects, and are not coronal in origin, consistent with previous observations and interpretations of SEP events connected westward of the associated flare.}


   \maketitle

\section{Introduction} 

While investigations into solar energetic particle (SEP) events have long been conducted (see review by \citealp{Reames2023}), recent advances in instrumentation have enabled new investigations into their characteristics. SEP events typically exhibit a velocity dispersion feature in the flux-by-energy spectrum, in which the initial release of a spectrum of particles near the Sun disperses as the population propagates to an observer. Although the initial population would have been released at nearly the same time, higher-energy (and thus higher-speed) particles will travel along the flux tube to the observer arriving before the lower-energy (lower-speed) particles, establishing a velocity dispersion in the onset times of SEPs. The fitting of this signature can then provide insight into both the time of initial release and the path length through which the particles traveled (e.g., \citealp{Zhao2019, Li2021, Wimmer2023}). 

One aspect in which newer instrumentation is advancing our ability to study SEP events is the increased energy resolution and dynamic range of energetic particle instruments on Solar Orbiter and Parker Solar Probe. This increase in energy resolution and range allows features to become more pronounced when shown in a higher-dimensional spectrogram form. Through this view, several SEP events have stood out as peculiar in that they have both a lower-energy velocity dispersion onset along with a higher-energy onset delay (e.g., \citealp{Cohen2024, Ding2025}). These events, in which higher-energy particles above some initial onset energy, $E_{0}$, become progressively more delayed with increasing energy, were initially termed ``inverse velocity dispersion'' events. While, the cause of this signature is strongly debated, it is not likely related to velocity dispersion effects. As such, in this study, they will be referred to by the key feature in the measurements, an observed delay in the maximum energy (DME) of the SEP event. 

Recent modeling work has investigated potential causes for DME events, suggesting that the maximum energy ($E_{max}$) of particles released by the portion of a coronal mass ejection (CME)-driven shock the observer is magnetically connected to may increase in time \citep{Ding2025}. This does not necessarily dictate whether the cause is due to changes in the CME-driven shock as the structure expands or simply that the observer becomes connected to a portion of the shock releasing a higher $E_{max}$ population, but it does explain how a delayed maximum energy event may be possible. Generally, these results are consistent with earlier modeling efforts to investigate both the importance of an observer's location relative to the shock front and the evolution of the shock structure \citep{Kallenrode1997}.

This importance of an observer's connection to an evolving shock structure to energy-dependent onsets may have some commonality with the cause of variations in SEP time profiles. For example, \citet{Cane1988} suggested several ``pathways" for observed SEP time profiles based on the longitude of the observer from the nose of the CME. In this framework, \citet{Cane1988} suggested that events magnetically connected near the associated flare site would observe particles accelerated from either the flare or coronal CME-driven shock initially, followed by relatively weaker fluxes from the continued acceleration at the CME-driven shock in interplanetary space. For observers with a connection to the Sun more westward of the flare site, the high flux component of energetic particles from the flare or coronal shock would no longer be observed, but instead particles from the CME-driven shock accelerated in interplanetary space would slowly intensify, resulting in an apparent ``delay" in the maximum intensity. This gradual rise in intensity was explained as the observer becoming increasingly well connected to stronger portions of the shock as the CME expanded \citep{Cane1988}.

Even earlier, \citet{McCracken1970} and \citet{Kahler1969} described a scenario in which observers connected westward of the flare site would have a delayed connection to SEPs dependent on connectivity to the expanding shock front, despite the assumption that the source of energetic ions was the original flare site. \citet{Reinhard1974} described observed delays between flare onset and peak SEP intensities, which increased with increasing longitudinal distance from the flare site, as being due to a combination of rapid coronal particle transport, large-scale drift processes, and diffusion. Further, \citet{VanHollebeke1975} reported that the delay between SEP-onset and peak intensity for 20-80 MeV protons systematically increased for events with solar connections increasingly westward from the associated flare site (i.e., more so eastward events when oriented on the central meridian), reaching tens of hours of delay. Their interpretation, again, relies on changes in the efficiency of particle acceleration at the point on the shock to which an observer is magnetically connected. As such, descriptions of ``delayed" SEP events (whether delay between flare and onset or onset and peak intensity) have long been discussed and have generally been explained through a combination of magnetic connectivity, particle transport, and CME expansion/evolution. It stands to be fully explored, however, how the occurrence and characteristics of DME SEP events may also depend on these aspects (e.g., relative observer location to the associated flare, connectivity to the CME-driven shock, etc).

This study aims to expand on our understanding of these DME events by conducting a statistical survey of Solar Orbiter observations to compare DME events to non-DME events, and to derive their key characteristics. Section 2 outlines the datasets and key methodologies used in the investigation, Section 3 presents the main results, and a discussion of the results and conclusions are in Section 4.

\section{Datasets and methodology} \label{sec:Method}

Solar Orbiter launched in February 2020 into an elliptical orbit around the Sun, with the perihelion being lowered periodically with Venus gravity assist maneuvers \citep{Muller2020}. Observations from the Solar Orbiter Energetic Particle Detector (EPD; \citealp{Rodriguez2020, Wimmer2021}) are used in this study to characterize SEP event onsets. Specifically, this study uses ion observations from the SupraThermal Electron and Proton sensor (STEP), Electron Proton Telescope (EPT), and High-Energy Telescope (HET) of EPD, which span the energy ranges of 6 – 80 keV, 25 keV – 6.9 MeV, and 6.8 MeV – 100 MeV, respectively. For this study, EPD data are averaged down to a 1 hr resolution to reduce the channel-to-channel deviation in onset times as determined by the fitting algorithm described below. Additionally, only observations from the sunward-pointed EPT and HET telescopes are used to better align with the field-of-view of STEP when combining the measurements \citep{Rodriguez2020, Wimmer2021}. Magnetic field measurements from the Solar Orbiter fluxgate vector magnetometer (MAG; \citealp{Horbury2020}) and plasma moments from the Proton and Alpha particle Sensor (PAS), part of the Solar Wind Analyser (SWA; \citealp{Owen2020}), are also used in this study. 

For all events, the associated flare site and CME speed were taken from the Community Coordinated Modeling Center (CCMC) Database Of Notifications, Knowledge, Information (DONKI\footnote{https://kauai.ccmc.gsfc.nasa.gov/DONKI/}) catalog. As discussed in \citet{Richardson2015}, parameters from DONKI are derived from observations from the Large Angle and Spectrometric Coronograph (LASCO, \citealp{Brueckner1995}) onboard the SOlar and Heliospheric Observatory (SOHO) and from the Sun Earth Connection Coronal and Heliospheric Investigation (SECCHI, \citealp{Howard2008}) coronagraph on Solar TErrestrial RElations Observatory (STEREO) incorporating triangulation, if available, using either the Stereoscopic CME Analysis Tool (SteroCAT, \citealp{Mays2015}) or the National Oceanic and Atmospheric Administration Space Weather Prediction Center CME Analysis Tool (SWPC-CAT, \citealp{Millward2013}). When triangulation is not available, a single spacecraft solution is used. The DONKI catalog provides a comprehensive set of CME characteristics computed in a more operational real-time fashion. While not as accurate as other reconstruction techniques, which are themselves known for their own associated uncertainties (e.g., \citealp{Key2020, Verbeke2023, Kay2024a}), the availability of the DONKI catalog has enabled many studies into CME characteristics and their impacts on the heliosphere (e.g., \citealp{Cavus2025}) and magnetosphere (e.g., \citealp{Weiler2025}), and space weather prediction validation (e.g., \citealp{Richardson2018}), along with various data-model and model-model comparisons (e.g., \citealp{Wold2018, Kay2024b, Valentino2024}). The CME speeds for 27 of the events are also included in the Coordinated Data Analysis Workshop (CDAW) catalog\footnote{https://cdaw.gsfc.nasa.gov/} and are comparable with the DONKI-provided CME speeds. There was on average an agreement to within $\sim$20\%, which is consistent with previous comparisons of CME speeds between CDAW and DONKI, showing those datasets to be generally well correlated (e.g., \citealp{Richardson2015}).

\begin{figure*}
    \centering
    \includegraphics[width=\linewidth]{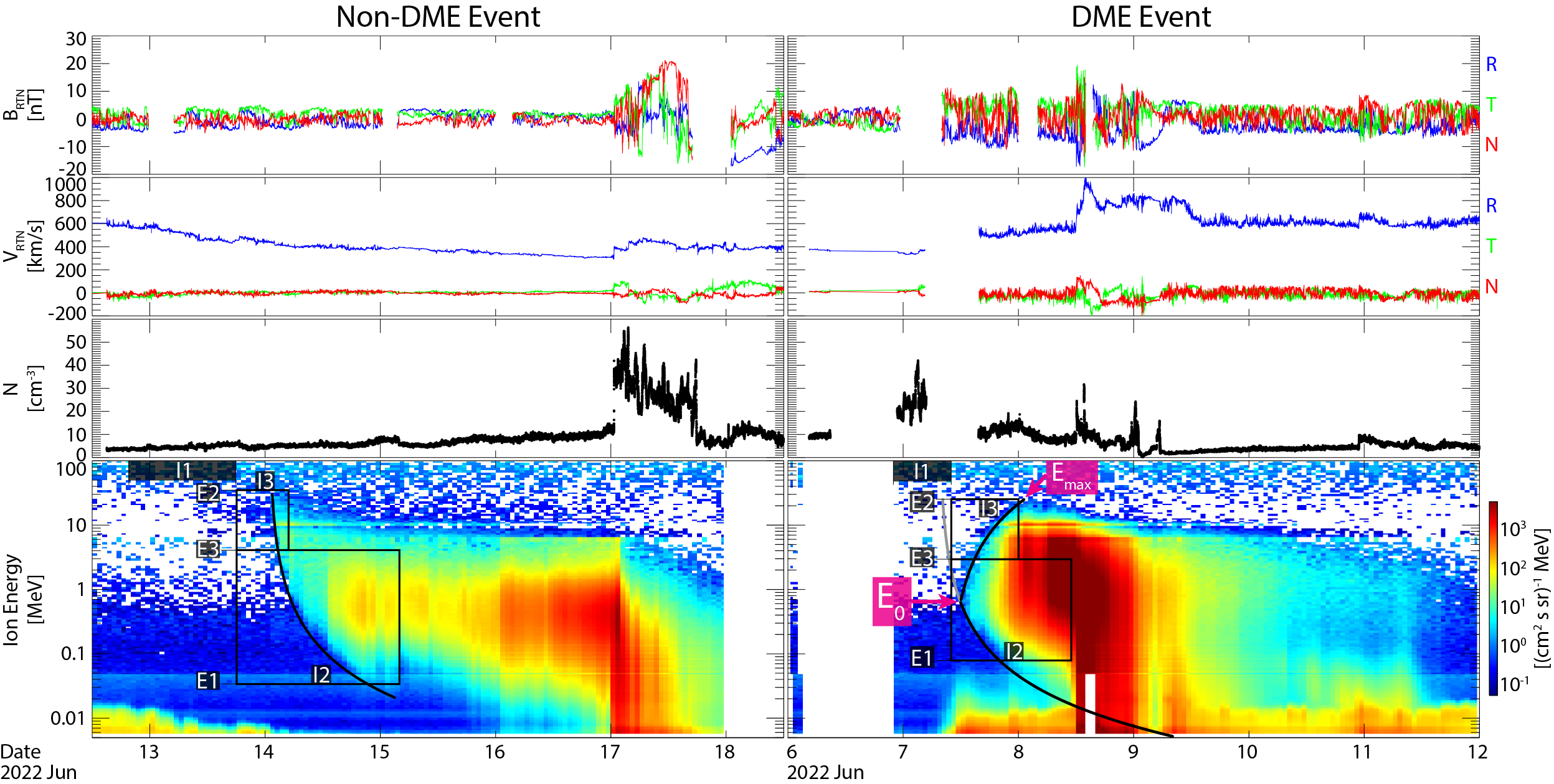}
    \caption{From top to bottom, magnetic field components from MAG in RTN coordinates, solar wind velocity components in RTN coordinates and density from SWA/PAS, and ion flux spectrograms from EPD (STEP/EPT/HET) for a non-DME event (left) and DME event (right). Black curves on the flux spectrograms denote the onset time fittings and the grey curve on the DME event flux spectrogram illustrates the extrapolated VDA fit for the higher energy portion. The point where the lower energy fit of the DME event changes to grey is $E_{0}$ (as labeled in magenta) and the end of the higher energy fits is $E_{max}$ (labeled for the DME event). By-eye identified intervals and energies used by the fitting algorithm are shown over the ion spectrogram for illustrative purposes (see text for description).}
    \label{fig:Fig1}
\end{figure*}

Four years (2020 July 01– 2024 July 01) of EPD observations were surveyed to identify 31 SEP events (shown in Table~\ref{tab:table1}) that met the following criteria: (1) SEP flux enhancements above background reaching up to at least 10 MeV, (2) isolated SEP onsets, not overlapping other enhancements or events, and (3) without ``drop-out" signatures (see for example \citealp{Ho2022, Wimmer2023}) occurring in the initial portion of the onset. The initial portion of the onset used for criterion~3 is defined as either the first six hours or prior to the maximum energy SEP onset, whichever is longer. To characterize the onsets of each of these events, three intervals were visually identified (Figure~\ref{fig:Fig1}): (I1) a representative background pre-event interval needing to last at least 12 hours before the SEP event, (I2) an interval of SEP enhancement from prior to the first arriving particles to after the arrival of the lowest-energy particles considered (E1, see below for definition) but before the decay phase of the lower-energy portion (energies between E1 and E3) of the SEP event, and (I3) an interval of SEP enhancement from just prior to the initial particle arrival to the time when the flux of the highest-energy particles begins to decay. Additionally, three energies were visually identified to be used by the fitting algorithm: (E1) the lowest energy channel to both see the SEP event and be above any background or other signals not associated with the SEP event, (E2) the highest energy in which an appreciable signal (i.e., above background) was observed, and (E3) the energy at which the fitting algorithm will switch from I2 at low energies to I3 at high energies. A transition between I2 and I3 is necessary because the time over which the event should be searched for high- and low-energy onsets differs between events. As long as E3 is chosen to be above the low-energy dispersive onset at the end of I3 and below the energies at which the higher-energy component begins to decay at the end of I3, its precise selection does not affect the fitting results. This is the case for both non-DME and DME events (Figure 1). 

\begin{table*}
\small
\caption{SEP event characteristics\label{tab:table1}}
\begin{tabular}{l c c c c c c c c c}
\hline
Event & Type & Onset Time$^{a}$ & $E_{max}$ & $E_{0}$ & $V_{sw}$ & CME Speed & SS to Flare$^{b}$ & SolO R$^{c}$ & $D_x-D_0$  \\
\# & & (UT) & (MeV) & (MeV) & (km/s) & (km/s) & (deg.) & (au) & (au) \\
\hline
1 & non-DME & 2020-11-24/14 & 21.24 & -- & 400$^{d}$ & 878 & -24.70 & 0.9 & -- \\
2 & non-DME & 2020-11-29/14 & 97.63 & -- & 400$^{d}$ & 1476 & -65.17 & 0.88 & -- \\
3 & non-DME & 2021-04-17/17 & 81.64 & -- & 377 & 828 & 62.91 & 0.84 & -- \\
4 & non-DME & 2021-05-07/21 & 97.63 & -- & 415 & 815 & 34.79 & 0.92 & -- \\
5 & non-DME & 2021-05-29/01 & 25.09 & -- & 301 & 824 & -80.32 & 0.95 & -- \\
6$^e$ & DME & 2021-08-26/20 & 23.12 & 4.27 & 326 & 599 & -8.16 & 0.62 & 0.025 \\
7 & non-DME & 2021-09-28/08 & 49.07 & -- & 392 & 649 & -31.89 & 0.62 & -- \\
8 & non-DME & 2021-11-09/18 & 49.07 & -- & 506 & 1310 & -72.87 & 0.88 & -- \\
9 & non-DME & 2022-01-18/19 & 27.20 & -- & 552 & 962 & -27.67 & 0.94 & -- \\
10 & non-DME & 2022-01-30/01 & 10.60 & -- & 374 & 628 & 46.64 & 0.86 & -- \\
11 & non-DME & 2022-03-21/05 & 97.63 & -- & 391 & 792 & -84.35 & 0.34 & -- \\
12 & non-DME & 2022-04-15/03 & 41.18 & -- & 547 & 521 & 1.94 & 0.50 & -- \\
13 & DME & 2022-06-07/12 & 25.09 & 0.63 &  380 & 836 & 44.37 & 0.96 & 0.339 \\
14 & non-DME & 2022-06-14/02 & 29.44 & -- & 634 & 830 & 14.50 & 0.98 & -- \\
15 & DME & 2022-06-26/10 & 12.43 & 0.90 & 412 & 950 & 70.51 & 1.01 & 0.345 \\
16 & DME & 2022-07-23/22 & 58.03 & 5.35 & 499 & 1582 & 24.16 & 0.99 & 0.070\\
17 & non-DME & 2022-11-19/14 & 25.09 & -- & 313 & 627 & 25.98 & 0.70 & -- \\
18 & DME & 2023-03-13/06 & 25.09 & 0.93 & 374 & 2127 & 172.34 & 0.61 & 0.409 \\
19$^f$ & DME & 2023-04-21/13 & 10.60 & 0.38 & 336 & 804 & 49.20 & 0.37 & 0.517 \\
20 & non-DME & 2023-05-16/19 & 81.64 & -- & 400$^{d}$ & 1480 & 71.87 & 0.69 & -- \\
21 & non-DME & 2023-09-11/22 & 97.63 & -- & 666 & 1716 & -6.05 & 0.58 & -- \\
22 & non-DME & 2023-09-24/19 & 41.18 & -- & 353 & 619 & 122.75 & 0.41 & -- \\
23 & non-DME & 2023-10-03/12 & 97.63 & -- & 358 & 632 & 14.45 & 0.31 & -- \\
24 & DME & 2023-11-02/12 & 49.07 & 1.75 & 278 & 699 & 50.37 & 0.57 & 0.302\\
25 & DME & 2023-11-09/16 & 53.38 & 0.68 & 730 & 782 & -2.18 & 0.66 & 0.047\\
26 & DME & 2023-12-24/22 & 29.44 & 2.22 & 347 & 790 & 47.77 & 0.94 & 0.215\\
27 & DME & 2023-12-31/14 & 75.11 & 2.82 & 291 & 671 & 12.33 & 0.95 & 0.627\\
28 & DME & 2024-01-21/02 & 19.65 & 1.02 & 400$^{d}$ & 978 & 61.77 & 0.92 & 0.149\\
29 & non-DME & 2024-01-29/06 & 81.64 & -- & 556 & 1888 & -10.75 & 0.90 & -- \\
30 & non-DME & 2024-05-27/09 & 68.97 & -- & 558 & 1254 & -21.01 & 0.83 & -- \\
31 & non-DME & 2024-06-22/00 & 68.97 & -- & 596 & 922 & -44.79 & 0.94 & -- \\
\hline
 \multicolumn{10}{l}{$^a$Onset time of first arriving particle.} \\
 \multicolumn{10}{l}{$^b$Angular separation between the Solar Orbiter source surface (SS = 2.5 Rs) footpoint and associated flare site.} \\
 \multicolumn{10}{l}{$^n$Radial distance of Solar Orbiter at the time of $E_0$.} \\
 \multicolumn{10}{l}{$^{d}$Assumed $V_{sw}$ due to absence of SWA/PAS moments.} \\
 \multicolumn{10}{l}{$^e$Very short lag-time between velocity dispersion profile and potential delayed signature ($\sim$2~hrs at $E_{max}$).} \\
 \multicolumn{10}{l}{$^f$Location of estimated $E_{max}$ release is beyond that of Solar Orbiter.} \\
\end{tabular}
\end{table*}

Using the average flux and standard deviations within the pre-event background interval (I1), a background flux was computed per energy channel per event. If no counts were recorded during the background interval for a specific energy channel, then a fixed value of 0.5 [(cm$^2$ s sr MeV)$^{-1}$] was used for the background flux and standard deviation for that channel. It should be noted that as long as this assumed background flux is sufficiently low, i.e., below the flux of SEP enhancement, then the choice of this background will not impact the results due to this value only being used for events and energy channels when the flux is below the detection level of the instrument. With this information, an algorithm was applied to search backwards in time from end of intervals I2 and I3 within the energy ranges below and above energy E3, respectively, for the first timestep where fluxes dropped below two standard deviations above the average background flux. 

To fit the per-energy channel onset times, a functional form is required. While typical velocity dispersion analysis (VDA) defines the time of onset as $\tau_{onset}=l / v_{p}$, where $l$ is the particle pathlength from where the particle spectra was released and $v_{p}$ is the particle speed, this functional form cannot describe the delayed arrival of higher-energy particles in DME events. Instead, the onset time for DME events would be expected to follow a piecewise equation such as,

\begin{equation}
    \tau_{onset}(E)=\left\{ \begin{array}{ll} \frac{l}{v_{p}(E)} \quad E \le E_{0} \\ \frac{l(E)}{v_{p}(E)} + \tau_{release}(E,...) \quad E > E_{0} \end{array} \right. 
\end{equation}

where $E$ is the particle energy, $E_0$ is the energy at which the onset times transition from the low energy velocity dispersive profile to the higher-energy delayed onset profile, and $\tau_{release}$ is the release time of a particle at a given energy within the higher-energy delayed range. The lower energy portion of this formulation (i.e., $E \le E_{0}$) is then identical to traditional VDA formalism and will capture the onset of typical velocity dispersion. For the higher-energy portion (i.e., $E > E_{0}$) of this equation, the pathlength of the particle now has an energy dependence due to the delay in release allowing the CME to expand closer to the observer, lowering the pathlength for higher energies. As $\tau_{release}$ may likely depend on many unknown quantities, and quantifying the change in particle pathlength in time as the CME-driven shock expands is non-trivial, this work instead introduces a fitting term $\alpha$ and fits the computed onset times to 

\begin{equation}
    \tau_{onset}(E)=\left\{ \begin{array}{ll} \frac{l}{v_{p}(E)} \quad E \le E_{0} \\ \alpha v_{p}(E) \quad E > E_{0} \end{array} \right. 
\end{equation}

This allows the piecewise fitting function to well describe both events with velocity dispersion at energies below $E_{0}$ and delayed onset times at energies above $E_{0}$ as well as events with only velocity dispersion (i.e., setting $E_{0} = E_{max}$). If the fitted solution to the DME portion (i.e., $E > E_{0}$) of the piecewise function at the observed $E_{max}$ of the SEP event (i.e., at the highest energy above background) was less than one hour from the estimate of the onset timing using only the lower-energy VDA portion of the equation ($E \le E_{0}$), then the event was re-fitted with only the VDA equation over all energies due to the delay being less than one temporal bin width, and the event is classified as a non-DME event. Conversely, if the delay between the fitted $E_{max}$ onset and the prediction from the lower energy VDA portion was greater than one hour then the event was classified as a DME event. Using this methodology, 11 of the 31 SEP events were identified as a DME event (see Table 1). All fits have been verified by eye and describe the onset times for the events well.

Two representative examples of applying this method are shown in Figure~\ref{fig:Fig1} for a typical velocity dispersion SEP event (left column) and a DME event (right column). The magnetic field (top row) and solar wind speed and density (middle row) are provided for context and are typical for isolated CME-associated SEP events. The EPD (STEP/EPT/HET) combined spectrograms are shown in the bottom row for both events. The intervals and energies used to determine $\tau_{onset}$ and identify DME or non-DME events are shown, along with the resulting fits. Although these input values are used in the fitting algorithm, the fitting algorithm is not highly sensitive to the exact values chosen. The lower energy range used (E1) was selected to avoid low-energy enhancements either during the pre-event background interval (e.g., in the non-DME event case in Figure 1) or during the event, but unrelated to the SEP onset (e.g., in the DME event case of Figure 1). It should be noted that the highest energy included (E2) could be a lower limit on the maximum energy reached due to instrument sensitivities. Although the fitting equation (Equation~2) is simplistic, it adequately captures the general profile of the delayed onsets for the analysis of this study for all DME events investigated.

As described in Section~1, previous studies have interpreted various characteristics of SEP onsets as being related to the relative location of an observer to the source of the SEPs (either the associated flare or CME-driven shock). So, to better understand these specific SEP events, the footpoints of Solar Orbiter relative to the associated flare site were estimated. This was computed through ballistic back mapping (e.g., see description in \citealp{Allen2023}) using the average solar wind speed observed over the duration of the SEP onset fit to a source surface (SS) at 2.5 Rs. If the solar wind speed was not available for a particular event (as noted in Table 1) then a nominal speed of 400 km/s was assumed. Although more advanced methods exist, this approach is sufficient for the objectives of this analysis, aiming at only a rough estimate of the magnetic connectivity of Solar Orbiter above the source surface to investigate large-scale trends.

\section{Results} \label{sec:results}

\begin{figure*}
    \centering
    \includegraphics[width=0.90\linewidth]{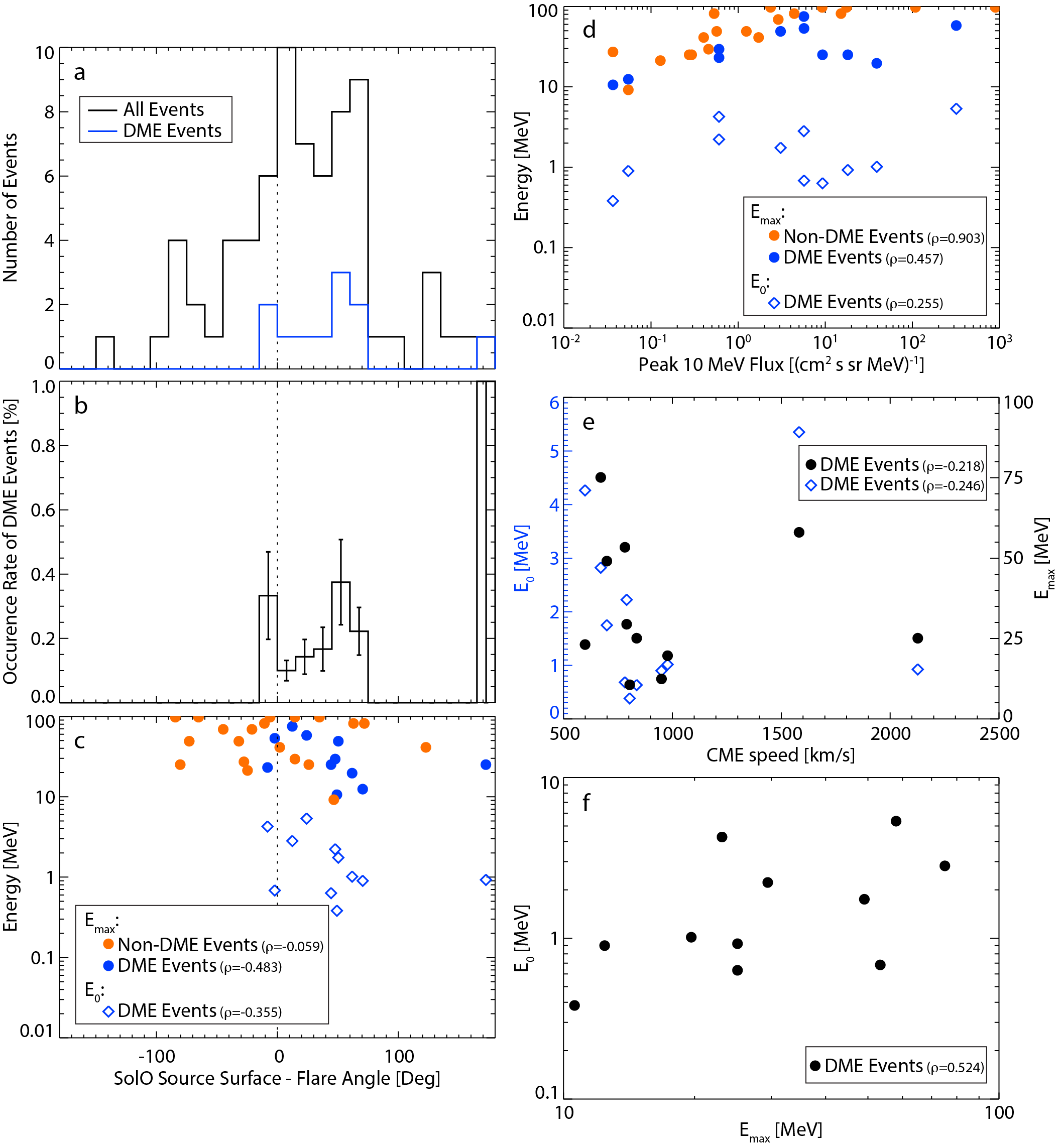}
    \caption{(a) Distribution of all SEP events (black) and DME events (blue) and (b) occurrence rate of DME events by foopoint location relative to associated flare site with associated Poisson error. $E_{max}$ for non-DME (orange circles) and DME (blue circles) events and DME event $E_0$ (blue diamonds) for (c) different footpoint locations and (d) different peak flux at 10~MeV. (e) $E_0$ (blue diamonds) and $E_{max}$ (black circles) values versus CME speed for DME events. (f) DME event $E_0$ as a function of $E_{max}$. Panels (c - f) include the Spearman's rank correlation coefficient within the legends provided.}
    \label{fig:Fig2}
\end{figure*}

To explore the spatial organization in the occurrence of DME SEP events, Figure~\ref{fig:Fig2}a shows the distribution of the longitudinal angle between the source surface footpoint location and the associated flare site for all 31 SEP events (black) and just the 11 DME SEP events (blue) as a histogram with bins 15 degrees in width. Of note, in this study the frame of reference is set to the associated flare longitude rather than a central meridian to investigate how magnetic connectivity may impact the observations. This is done to investigate the hypothesis that the spacecraft footpoint relative to the flare location may be a more physically significant ordering factor than the spacecraft location. Using this reference frame results in the axes of Figure~\ref{fig:Fig2}a-c to be reversed from earlier studies that ordered observations based on the central median (e.g., \citealp{Kahler1969, McCracken1970, VanHollebeke1975, Cane1988}), so care must be taken when comparing to these other studies. In this frame, positive (negative) angles indicate footpoint locations that are westward (eastward) of the flare site. While Figure~\ref{fig:Fig2}a illustrates the raw numbers of events per footpoint offset, an occurrence ratio (in this case, the number of DME events divided by the total number of SEP events for a given bin in angle space) is an often used and more robust statistical measure of how the percent of events that are DME vary and mitigates observational biases. Taking this ratio of DME events to all events, Figure~\ref{fig:Fig2}b shows that the occurrence rate of DME events have a clear bias to the westward side of the flare site. While the statistics are limited, the bias towards the westward side of the flare site is pronounced. 

In addition, comparing DME and non-DME events, the $E_{max}$ of both types of events and the value of $E_{0}$ for DME events are shown in Figure~\ref{fig:Fig2}c. To assess the correlation between these values and the angle between the estimated footpoint and flare site, Spearman's ($\rho$) rank correlation coefficients are provided in Figure~\ref{fig:Fig2}c. This metric was chosen over the typical Pearson correlation coefficient due to the small number of events under investigation. There is negligible correlation between the non-DME $E_{max}$ and the footpoint location relative to the associated flare ($\rho = -0.059$), with only moderate correlation for DME event $E_{0}$ ($\rho = -0.483$) and a slightly weaker correlation for DME event $E_{max}$ ($\rho = -0.355$). However, several non-DME events have $E_{max}$ near 100 MeV which is also the maximum observable energy, allowing the possibility that some of the 100 MeV non-DME events would be classified differently if higher-energy (i.e., $>~100$~MeV) fluxes were observable. With this upper energy limit of the observations, there is no clear significant difference in the maximum energy between the two types of events within the energy range of EPD. It should be noted, however, that Figure~\ref{fig:Fig2}c illustrates variations between different events rather than within a single event (i.e., spatiotemporal evolution), which would require multi-spacecraft analysis (e.g., \citealp{Lario2013, Kollhoff2021, Mason2021}). As such, this analysis should not be construed as indicating that any of these parameters, such as $E_{0}$, would not vary spatially for a single event. Instead, this only means that there is not a strong dependence across all included events (that is, while a single CME-associated shock may result in a longitudinally varied $E_{0}$ such as in \citet{Ding2025}, the variability of $E_{0}$ between events may be greater than the spatial variations in $E_{0}$ for a single event, resulting in this dependence not being seen across the sampling of events in this study).

Figure~\ref{fig:Fig2}d explores the relationship between $E_{max}$ and $E_{0}$ compared to the peak 10~MeV flux for each event. The choice of 10~MeV as the reference flux was made because it is the highest energy required for the SEP enhancement to be included in our event list. Unsurprisingly, as the peak flux increases, the maximum energy of observed particle enhancement also increases for both DME and non-DME events (with a very strong Spearman's rank correlation for non-DME events, $\rho=0.903$, and a moderate correlation for DME events, $\rho = 0.457$). When comparing the two classifications of events, DME events are seen to have a slightly lower maximum energy for the same peak 10~MeV flux. Similarly, the value of $E_{0}$ seems to slightly increase with increasing peak 10~MeV flux (with a weak correlation of $\rho = 0.255$), however more events are required to constrain the statistical significance of these trends. One potential explanation for the difference in $E_{max}$ between the types of events, particularly at higher peak fluxes, may potentially be indicative of the break energies of non-DME events being higher than for DME events. This would shift the exponential cut-off between the two event types, despite similar peak 10~MeV flux values. Further investigation into this possibility will be the included in future studies.

As the CME speed can impact the strength of the driven shock, which in turn impacts the peak SEP flux (e.g., \citealp{Park2017}), it may potentially have implications for $E_{0}$ or $E_{max}$. To investigate this, Figure~\ref{fig:Fig2}e shows both $E_{0}$ (black) and $E_{max}$ (blue) as a function of CME speed for DME events. Although most of the DME events have been observed for events with slower CME speeds (i.e., less than 1000 km/s), there is no clear correlation in CME speed and either $E_{0}$ or $E_{max}$ ($\rho = -0.246$ and $\rho = -0.218$, respectively). This suggests that CME speed alone is not the dominant factor for determining $E_{0}$ or $E_{max}$. 

The relationship between $E_{0}$ and $E_{max}$ is investigated in Figure~\ref{fig:Fig2}f. While there exists significant scatter, a moderate correlation is found ($\rho = 0.5424$). However, the large degree of scatter in the points does suggest that this is a weak relationship and that the Spearman's rank correlation coefficient may decrease with additional events. 

Together, the results of Figure~\ref{fig:Fig2} suggest that while most of the DME events have footpoints westward of the flare site, the relative values of $E_{0}$ and $E_{max}$ are not solely set by footpoint location or the CME speed, showing only a weak to moderate correlation to either quantity by itself. As such, these characteristic quantities are most likely determined by additional factors, such as the shock geometry and characteristics at the point of magnetic connectivity, rather than only the footpoint location or CME speed alone.

\begin{figure*}
    \centering
    \includegraphics[width=\linewidth]{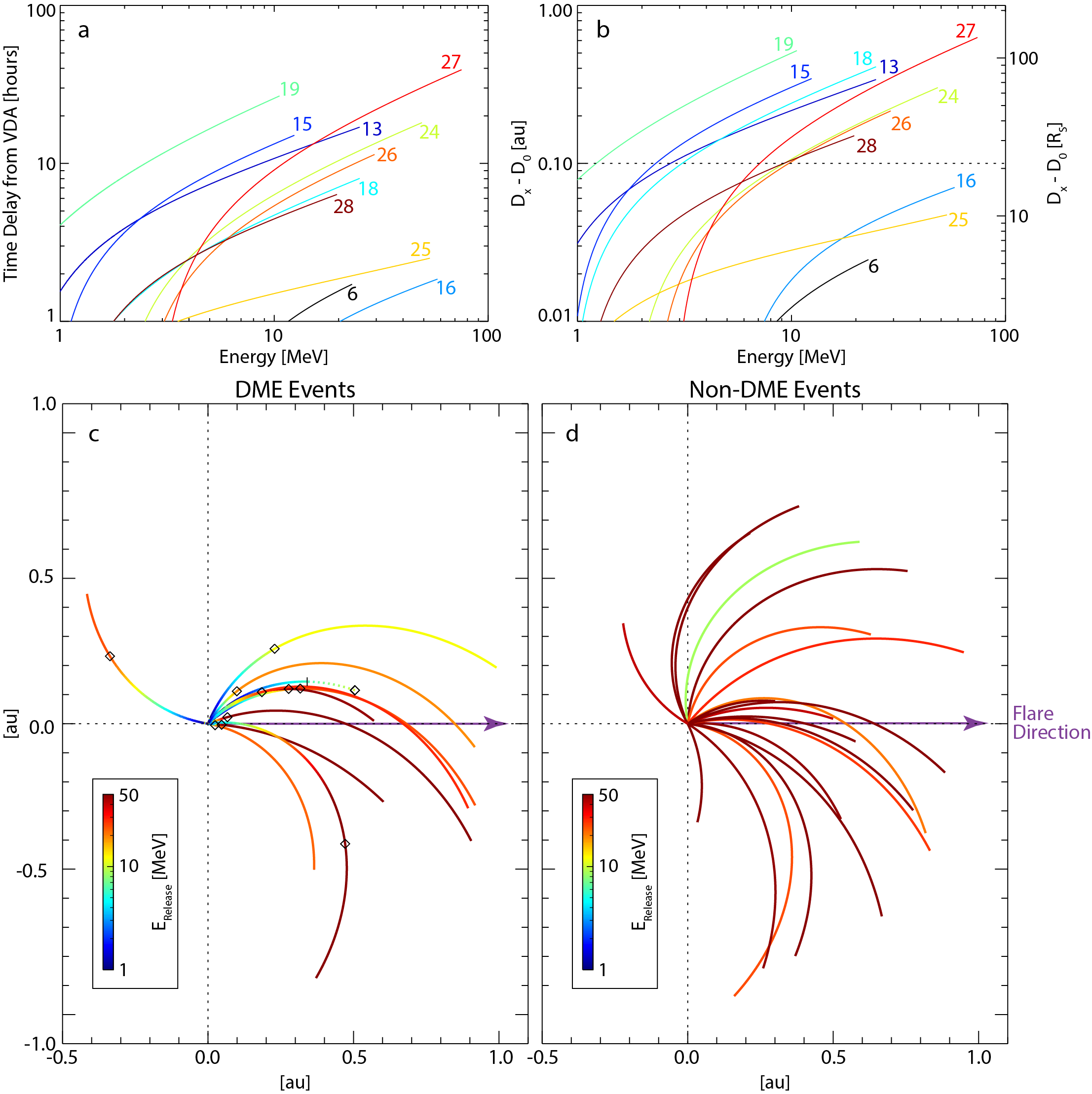}
    \caption{(a) Time delay between lower-energy VDA extrapolated to higher energies of DME events and the fitted higher-energy arrival times by energy. (b) Distance from initial release of particles and higher-energy release of particles versus particle energy for DME events. Numbers in panels a and b indicate the event number for each DME event listed in Table~\ref{tab:table1}. Idealized Parker spiral magnetic fieldlines colored by the maximum energy of particles being released at that heliospheric distance ($D_x$), neglecting the height of initial particle release ($D_0$), for (c) DME events and (d) non-DME events. Diamonds in panel c denote the location of $E_{max}$ release for the event, with the length of each Parker spiral fieldline extending to the location of Solar Orbiter for a given event (solid line). The dashed spiral in panel c extends from the location of Solar Orbiter to the computed location of maximum release energy for event 19. Panels c and d are oriented such that the location of the associated flare site would be directed along the positive x-axis, denoted by the purple arrow.}
    \label{fig:Fig3}
\end{figure*}

A fundamental aspect of DME events is the delay between when the particles would have been expected if released at the same time as the lower-energy component and when they were actually observed by Solar Orbiter. To better explore this, the low energy velocity dispersion onset fit was extended to higher energies (grey curve in Figure~\ref{fig:Fig1}, right). From this, each energy channel between $E_{0}$ and $E_{max}$ was used to compute a delay time from the two fitted curves (i.e., extended low-energy velocity dispersion and fitted higher-energy onset). These delays are shown as a function of energy in Figure~\ref{fig:Fig3}a. Using the CME speed as a crude estimate of CME propagation, this delay can then be converted into a distance between higher-energy release ($D_x$) and the original low-energy component release ($D_0$) as a function of energy (Figure~\ref{fig:Fig3}b). As the location of original particle release (i.e., flare site versus coronal CME, see \citealp{Li2023}) is not known, nor estimated here, this is only presented as a distance between the initial release of SEPs and first release of the higher energy population at a given energy. Since this approach uses a uniform CME speed estimated from DONKI, uncertainties exist in the actual CME speed, and more specifically the speed of the CME at the point Solar Orbiter was magnetically connected to; however, this initial approach provides significant insight into DME SEP events. For most of the DME events, the distance between the initial release and the release of few to 10’s MeV particles is over 0.1 au, or greater than $\sim$20~Rs, suggesting that this release is not a coronal process but one occurring within the inner heliosphere. 

Using these estimated distances and measured solar wind speeds during the SEP onset, a view of where the initial particle release would occur for different energies during DME events is shown (Figure~\ref{fig:Fig3}c) in comparison to a similar display of idealized Parker spiral magnetic fieldlines colored by their initial $E_{max}$ for non-DME events (Figure~\ref{fig:Fig3}d). Each solid Parker spiral extends out to the location of Solar Orbiter for the given event and the curvature of the spiral is determined by the observed solar wind speed when available. For one DME event (event 19 in Table~\ref{tab:table1}), the estimated distance of event $E_{max}$ release was further out in heliocentric distance than the location of Solar Orbiter, indicated by a dashed curve in Figure~\ref{fig:Fig3}c. Further analysis of this specific event would be required to better explore the cause of this in detail, but the general conclusions of this analysis are not changed with either the inclusion or exclusion of this event and the general characteristics follow the trends of the other DME events. There is no clear dependence on the radial location of Solar Orbiter and detection of either DME or non-DME type events, suggesting that the classification is not biased to a certain radial location of Solar Orbiter. 

Notably, many of the DME events were observed by Solar Orbiter at larger inner heliospheric distances (i.e., larger than 0.5~au, see Table~\ref{tab:table1}), rather than being isolated to regions near perihelion, such as the Parker Solar Probe observations in \citet{Cohen2024}. Diamonds indicate where the $E_{max}$ release for a given SEP event would have occurred along the Solar Orbiter-connected flux tube for DME events. It should be noted that the offset from the Sun for the location of initial release is not included in this schematic but would be estimated at being less than $\sim$10 Rs for either flare or coronal CME release \citep{Li2023}. Figure~\ref{fig:Fig3}c shows that DME events have a rapid increase in the energies being released into the observed flux tube in the westward region from the flare site (i.e., a strong gradient in $E_{release}$ with respect to heliospheric distance). While one event has an $E_{max}$ release in the eastward quadrant, it has a notably slower increase in energy of released particles. Comparatively, the non-DME events have their maximum release energy at the initial release location and as such only shows the range of observed $E_{max}$ values (Figure~\ref{fig:Fig3}d). We note that this study does not attempt to investigate relaxation or inhibition of the maximum release energy after the point of maximum release energy (i.e., this study investigates the increase in $E_{release}$ with respect to heliospheric distance up to the point of $E_{max}$ for an event and does not investigate how the maximum particle energy released into a flux tube changed beyond that point), as this will be the focus of future work. As such, the color illustrated beyond the locations of $E_{max}$ release (diamonds in Figure~\ref{fig:Fig3}c) only indicates the maximum energy of particles that have been released into that flux tube by a given distance, rather than specifically at that location.

\begin{figure}
    \centering
    \includegraphics[width=0.96\linewidth]{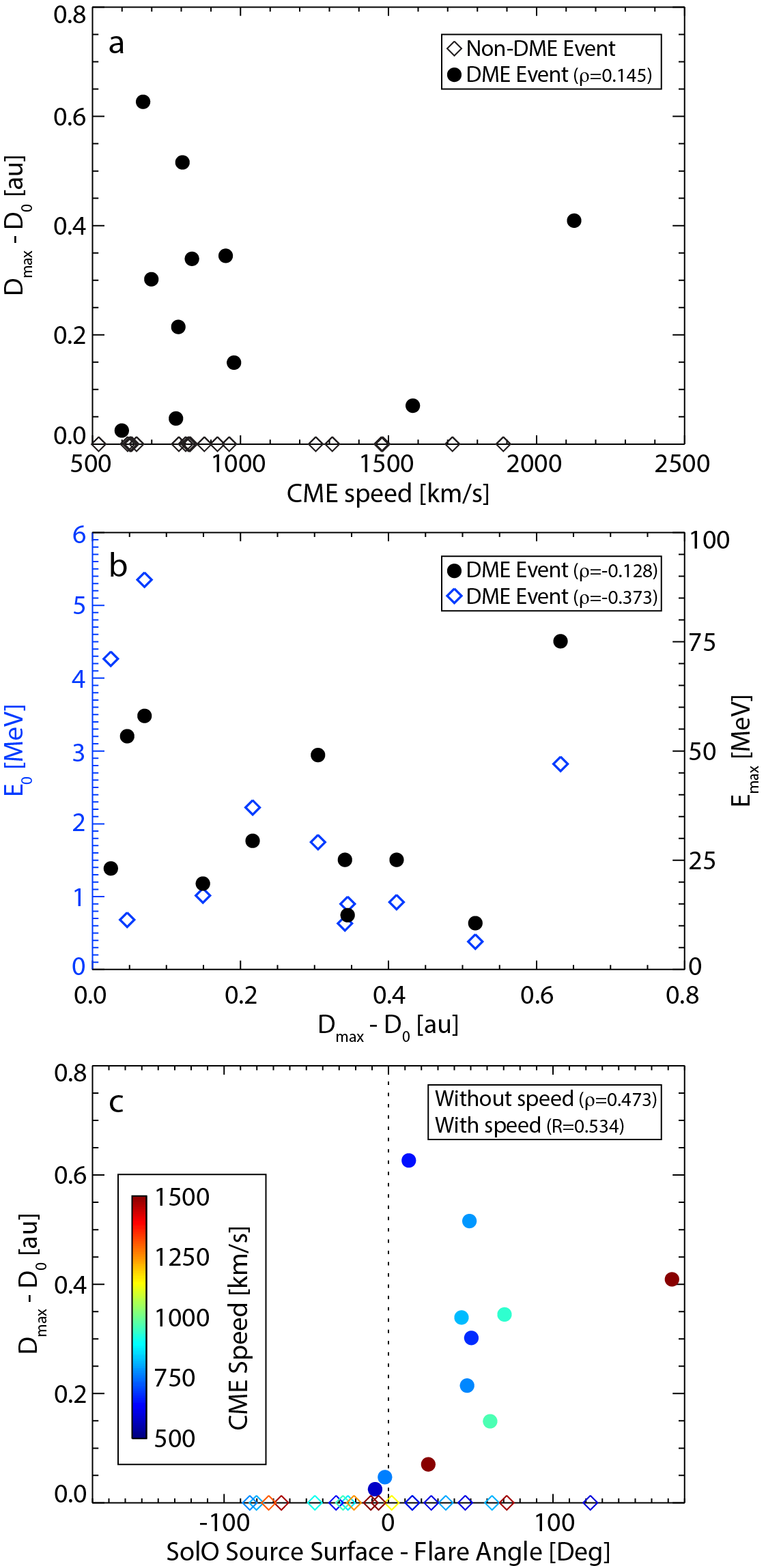}
    \caption{(a) Distance between initial particle release and the eventual point of maximum energy particle release into the flux tube observed by Solar Orbiter versus CME speed for DME events (solid circles) and non-DME events (open diamonds). (b) $E_0$ and $E_{max}$ versus the distance between initial and maximum energy release. (c) Distance between initial and maximum energy release versus footpoint location relative to the associated flare site for different CME speeds (indicated by symbol color).}
    \label{fig:Fig4}
\end{figure}

Further exploring the potential dependencies of the distance between initial release (at $D_0$) and final $E_{max}$ particle release into the flux tube (at $D_{max}$), this distance is shown versus CME speed (Figure~\ref{fig:Fig4}a). No clear correlation is found ($\rho = 0.145$), and notably the range of CME speeds spanned by the DME events also correspond to a non-negligible number of non-DME events. Given the lack of a significant trend, the CME speed is then not likely the critical criterion for DME events and the speed itself does not seem to be a critical factor in determining event type. Additionally, $E_{0}$ and $E_{max}$ do not have a strong correlation to the distance between initial and maximum energy release (Figure~\ref{fig:Fig4}b, $\rho = -0.373$ and $\rho = -0.128$, respectively), indicating that the distance itself is not, to zeroth-order, what establishes either the $E_{0}$ or $E_{max}$ values. While not significant, the weak association between $E_{0}$ and the distance between initial and maximum energy release may in part be due to large $E_0$ occurring only for events with an initial shock connection point that is already an efficient accelerator, thereby leaving less possibility for the connection to either change to a more efficient region, or for the shock itself to become significantly more efficient with time. This may then lessen the distance over which the CME can expand while still providing ever higher energy particles to the flux tube the observer is on. 

However, some correlation is observed between the distance between release locations and the location of the Solar Orbiter footpoints, especially when accounting for CME speed (Figure~\ref{fig:Fig4}c). From Figure~\ref{fig:Fig4}c, events with magnetic connection further westward from the associated flare site correspond to a larger distance between initial release of particles and the release of particles with the maximum energy observed for the SEP event ($\rho = 0.473$). Within a given range of angular separations, faster CMEs may have lower distances between initial and highest energy particle releases, although more events are needed to investigate this with more statistical significance. Computing a multi-variable Pearson correlation coefficient ($R$), the correlation between $D_{max} - D_0$ and both the footpoint-flare angle and CME speed indicates a moderate association ($R=0.534$). The largest separation point, which appears as a potential outlier to this trend, corresponds to the event in which the energy of release only gradually increased with distance. As such, it is possible and probable that additional impacts, such as changes to the CME with propagation, could be occurring beyond the general trend of events. Additional events are required to better constrain this relationship and gain further insight.

\section{Discussion and conclusions}

A survey of SEP events observed by Solar Orbiter was conducted in which 31 isolated SEP events with appreciable flux and energies reaching at least 10~MeV were investigated. Of these events, 11 had a delayed maximum energy from a velocity dispersion profile, referred to here as delayed maximum energy (DME) events. While some previous studies have referred to these as ''inverse velocity dispersion" events, there is no indication of velocity dispersion playing any role in this delayed arrival signature, and so this name may be misleading and attribute a physical cause not occurring in these events. 

There is a notable bias in the occurrence of these events to have source surface connection on the westward side of the associated flare site. However, there is no clear association between $E_{max}$ or $E_{0}$ (initial onset energy) and the separation between Solar Orbiter footpoint and the associated flare site. As such, while these DME events appear to occur due to changes in release times of particles into flux tubes on the westward side of the flare site, the angular separation itself does not set the associated energies. However, it is unclear if the absence of a trend is due to larger event-to-event variability than single-event spatial variability. Multi-spacecraft analysis is required to better understand how these properties change for a single event, but across many events, these values are not organized well by relative footpoint location.

Further, the transition between the velocity dispersion profile and the delayed onset energies does not appear to have a dependence on CME speed. While the Solar Orbiter DME events are generally observed for slower CMEs (less than 1000 km/s), this may be a bias in CMEs observed in the Solar Orbiter era thus far, as many of the non-DME events also occurred for slow CMEs. This lack of dependence between $E_{0}$ and CME speed then indicates that CME speed is not the zeroth-order driver of these events. This is not surprising, however, as the more relevant quantities for particle acceleration would depend on the Alfv{\'e}nic Mach number and compression ratios of the CME-driven shock, rather than the shock speed alone. As such, further investigation of the CME-driven shock kinematics and characteristics at the points connected with the observer are required.

Computing the time delay between when the particles would have been observed if arriving from the initial particle release and the fitted higher energy delayed profile allows for a characterization of this delay time by particle speed (or energy). The resulting estimate of CME propagation from the initial particle release to the observed higher-energy particles was typically over 0.1 au (or over $\sim$20 Rs) for particles in the few to 10’s~MeV in most DME events, with some extending out to over 0.5~au. Notably, assuming a highly improbable slower CME speed at the connection point than that estimated by DONKI would still result in a release well within the inner heliosphere (e.g., a 50\% slower CME would result in higher-energy releases 0.25~au further out than the initial particle release). This indicates that the delayed release is due to an inner heliospheric process, rather than a coronal process. 

This analysis does not include higher-order impacts, such as the delayed release particles having a shorter pathlength to the observer than the initially released populations (due to particles above $E_0$ originating from further out in the heliosphere) and that the distance of the initial release of particles would be to some finite height in the corona. Additionally, impacts from the increase in parallel mean free path with increasing distance from the Sun can result in additional scattering for particles released closer to the Sun than the delayed particles released further from the Sun (e.g., \citealp{He2012}). The combined result of these competing terms, i.e., the combination of shorter pathlengths than those assumed here but with less scattering for the delayed particles than the earlier released $E<E_0$ particles, is difficult to assess but would likely not typically be as significant as the delay in particle release or from CME expansion itself. As such, the general trends shown here provide a rough estimate of the spatial variability in the release of particles into flux tubes connected to Solar Orbiter.

The distance between the initial particle release and the release of the maximum energy particles for the event ($D_{max} - D_0$) does not appear to be well correlated with CME speed alone (Figure~\ref{fig:Fig4}a), and $E_{max}$ does not show strong signs of direct correlation with $D_{max} - D_0$ (Figure~\ref{fig:Fig4}b). While there is a weak association between $E_0$ and the difference in release heights, this could be due to events with relatively high $E_0$ being initially connected to a portion of the shock that is already efficiently accelerating particles, thereby limiting the distance the CME could expand to continuously release populations with ever increasing energy on the flux tube. However, a trend between $D_{max} - D_0$ and footpoint location relative to the associated flare site is observed, in that the distances appear to increase with increasingly westward footpoint locations (Figure~\ref{fig:Fig4}c). With a possible correlation with increasing CME speeds for larger distances and further westward footpoint locations, this trend also shows evidence of a potential secondary impact of CME speed, although more events are needed to investigate this possibility. Combined, these observations suggest that DME event flux tubes connected to the inner heliospheric region westward of an associated flare site have continued release of increasingly higher-energy particles as the CME-driven shock expands outward to potentially 0.5 au. 

This picture then is in remarkably good agreement with recent modeling work explaining these DME events as being due to changes in the maximum energy of particles being released into the flux tube from an expanding CME-driven shock \citep{Ding2025}. Additionally, this explanation of the cause of DME events is similar to that given for early statistical studies of SEP events focused on SEP time-profiles, in that both relate to the connectivity of an observer to an expanding shock (e.g., \citealp{Kahler1969, McCracken1970, VanHollebeke1975, Cane1988}). For example, the cause of events with “delayed” intensity peaks determined by \citet{Cane1988} was also attributed to the observer being connected westward of the associated flare location and thereby having a magnetic connection to the stronger portions of the expanding CME-driven shock later in the event, as compared to short-lived eastward events. While these previous works did not directly describe DME-type signatures like those observed by Solar Orbiter, this may be the result of these studies focusing on single channel or limited few channel, higher-energy (i.e., above 1 MeV, which is found here to be the nominal $E_{0}$ value) flux (or count-rate) time-series observations. This would be similar to taking only a narrow energy-slice of a DME event spectrogram in the delayed energy range (i.e., taking a single energy slice above $E_{0}$ in Figure 1) and comparing it to when the particles would have been expected from the flare onset. Therefore, all of these studies are possibly describing different aspects of a portion of the same root phenomena as observed here. The expanded energy resolution and range of Solar Orbiter EPD now enables for more robust exploration into the causes of these types of events to better elucidate particle acceleration, release, and transport in the inner heliosphere. This expanded energy range of EPD may also have allowed the events to be more prominent in the observations, which motivates re-analysis of older datasets for signs of DME-type signatures that may not have been as well noticed previously.

Beyond the aspect of time-varying magnetic connection to the shock impacting the release of of particles into the flux tube inhabited by the observer, acceleration at the shock is also time-varying. As such particles have more time to be accelerated to higher energies through crossing the shock front multiple times as the shock propagates outwards from the Sun (e.g., first-order Fermi acceleration, \citealp{Fermi1954}). Additionally, the acceleration time of particles and subsequent propagation to an observer are highly influenced by the diffusion coefficient in front of the shock (see \citealp{Drury1983}). As such, DME events are likely due to a combination of an observer's evolving magnetic connectivity and time-varying particle acceleration at CME-driven shock fronts as they expand into the upstream, structured solar wind. Better understanding the relative role of these different processes will require additional analysis and modeling work, and this class of events may provide a means to better explore this parameter space.

While the DME SEP events observed on Solar Orbiter tend to last for multiple hours (as seen in Figure 1), studies using Parker Solar Probe have recently reported observations of similar features on the order of minutes very close to the Sun (e.g., \citealp{Cohen2024, Kouloumvakos2025}). The interpretation of the Parker Solar Probe observations has been that the signature is due to sampling of a CME-associated shock as it is still accelerating particles to higher energies. While this may be the case for events very near the Sun (i.e., within 15 Rs), it is hard to justify this effect being able to create delay times on the orders of hours to days as seen further out in the heliosphere given the typical speeds of CME propagation. Further, this does not explain the spatial bias and characteristics of DME events observed in this study. It is then likely that the very near-Sun DME-type of events observed with Parker Solar Probe are probing coronal shock processes, whereas the longer delay events observed by Solar Orbiter are more largely impacted by inner heliospheric-driven changes to the maximum energy of particles released into flux tubes as the CME expands. This motivates future investigations into both regimes of processes with the understanding that they are likely to inform us of different types of fundamental processes occurring in the corona and the inner heliosphere.

This initial investigation also sets the stage for a more in-depth analysis of these types of events across multiple missions (e.g., Solar Orbiter and Parker Solar Probe). Future work will focus on expanding the data set for analysis as more observations are collected and incorporating relaxation of $E_{max}$ back to lower energy as the CME expands outward, particularly for eastward events. Additionally, future work will also utilize observations from Solar Orbiter EPD/Suprathermal Ion Spectrograph to better understand the mass and mass-per-charge dependencies of both particle release and transport for these DME-type events, such as those suggested in \citet{Ding2025}. Additional CME complexities, such as CME width, lateral expansion, and simulated evolution of shock parameters may also provide an important context for these events. Lastly, in-depth modeling and event reconstruction is required to better characterize shock dynamics occurring at the point to which an observer is connected, and how this varies in time, to better refine our understanding of the dominant drivers responsible for these observations.

\begin{acknowledgements}
Solar Orbiter is a mission of international cooperation between ESA and NASA, operated by ESA. The Suprathermal Ion Spectrograph (SIS) is a European facility instrument funded by ESA during construction, and NASA funds the instrument during Phase E operation. Solar Orbiter post-launch work at SwRI is supported by NASA contract 80GFSC25CA035, and we thank NASA headquarters and the NASA/GSFC Solar Orbiter project office for their continuing support. JRP acknowledges the financial support by the Spanish Ministerio de Ciencia, Innovaci\'on y Universidades FEDER/MCIU/AEI Projects ESP2017-88436-R and PID2019- 104863RB-I00/AEI/10.13039/501100011033. RFWS thanks ESA for supporting the build of SIS under contract number SOL.ASTR.CON.00004; the German Federal Ministry for Economic Affairs and Energy and the German Space Agency (Deutsches Zentrum f\"ur Luft- und Raumfahrt, e.V., (DLR)) for their unwavering support under grant numbers 50OT0901, 50OT1202, 50OT1702, and 50OT2002; and the University of Kiel and the Land Schleswig-Holstein for their support of SIS. RCA acknowledges support from NASA grant 80NSSC21K0733. RCA and RJF additionally thank NASA grant 80NSSC24K0908. RCA and SKV acknowledge support from NASA grant 80NSSC22K0993. We acknowledge the Community Coordinated Modeling Center (CCMC) at Goddard Space Flight Center for the use of the DONKI catalog.
\end{acknowledgements}

\bibliography{sample7}{}
\bibliographystyle{aa}

\end{document}